\documentclass[
    ,final            
  ]
  {aipproc}

\layoutstyle{6x9}

\begin{document}

\title{A Review of Coronagraphic Observations of Shocks Driven by
  Coronal Mass Ejections} 


\author{Angelos Vourlidas}{ 
    address={Code 7663, Naval Research Laboratory,Washington, DC 20375, USA},
}
 \author{Veronica Ontiveros}{
   address={Instituto de Geofisica, Universidad Nacional Autonoma de
   Mexico, DF, 04510, MEXICO},
 altaddress={CEOSR, George Mason University, Fairfax VA, 22030, USA}
 }

\begin{abstract}
  The existence of shocks driven by Coronal Mass Ejections (CMEs) has
  always been assumed based on the superalfvenic speeds for some of
  these events and on indirect evidence such as radio bursts and
  distant streamer deflections. However, the direct signature of the
  plasma enhancement at the shock front has escaped detection until
  recently. Since 2003, work on LASCO observations has shown that
  CME-driven shocks can be detected by white light coronagraph
  observations from a few solar radii to at least 20 R$_{sun}$. Shock
  properties, such as the density compression ratio and their
  direction can be extracted from the data. We review this work here
  and demonstrate how to recognize the various shock morphologies in
  the images. We also discuss how the two-viewpoint coronagraph
  observations from the STEREO mission allow the reconstruction of
  the 3D envelope of the shock revealing some interesting properties
  of the shocks (e.g., anisotropic expansion).
\end{abstract}

\maketitle


\section{The Long Search for CME-driven Shocks}

Ever since the accumulation of the first statistics on the properties
of CMEs in the 1970s \cite{gosling76}, it was
realized that many CMEs propagate at speeds in excess of 1000 km/s at
a few R$_{sun}$ above the solar surface which are higher than the
Alfv\'{e}n speed at these heights ($\sim800$
km/s; \cite{mann99}). Therefore, it is expected that CMEs drive
shocks in the corona and that the density compression at the shock front
could lead to detectable signatures in coronagraph images.  

The search for such signatures started with the \textit{Skylab\/}
images and \citep{hildner75, dulk76} where the first ones to
report the existence of a faint front ahead of the main CME
interpreting it as evidence of a bow shock. These so-called
'forerunners' were further analyzed by \cite{jackson78} and
\cite{jackson81}. However, \cite{karpen87} demonstrated that, in the
higher quality \textit{Solwind\/} images, these features were part of
the transient itself and were not associated solely or
even consistently with fast events. Others suggested that the bright
loop-like front of some transients was the enhancement from a fast MHD
shock \cite{maxwell81, stein85} but it was quickly pointed out by
\cite{sime84} that slow events also exhibit such loop-like fronts. The
works by \citep{sime84} and \citep{karpen87} casted  doubt
on whether the density enhancement from the shock could ever be
detected directly in the images. Indeed, most published CME shock
'detections' relied on indirect evidence such as distance streamer
deflections \cite{gosling74, michels84, sheeley00} or deductive
reasoning (e.g., fast lateral expansion, no streamer motion before the
CME) \cite{sime87}. 

While these analyses strongly supported the existence of CME-driven
shocks, they did not provide unambiguous detection of shock signatures
in coronagraph images. This was not a satisfactory
situation. CME-driven shocks are the main (if not the only)
accelerators of relativistic particles, the so-called solar energetic
particle (SEP) events, which have major space weather
implications. Many studies have indicated that the acceleration of
SEPs occurs between 4-10 R$_{sun}$ when the CME shock is at the
coronagraph field of view. Furthermore, CME shocks are the main means
of interaction between successive CMEs and may increase the
geoeffectiveness of such events \cite{lugaz05}. It is, therefore,
important to be able to follow their evolution in the inner corona and compare observations with the sophisticated simulations currently available \cite{liu08} .It is
obvious that from a space weather and SEP analysis standpoints the
direct detection and measurement of the parameters of CME driven
shocks is critical. It is fortunate that the recent years has brought
major advances in both of these issues.

The first direct detection of the density enhancement from a
CME-driven shock was reported by \cite{vou03} based on calibrated
LASCO images. The authors used an MHD model of the event to verify
that a shock was indeed expected at the observed location. They also
provided the first direct connection between a shock and the
associated streamer deflection and presented a few more examples of
such shocks. This enabled \cite{yan06} to recognize the density
compression from the shock at the CME flanks and associate it with a
type-III burst. More recently, \cite{ontiveros09} showed that faint
fronts can be detected ahead of the majority of fast CMEs, they
derived the density compression ratio at the shock front and
demonstrated that the observed density profiles are consistent with
line-of-sight (LOS) integration through a bowshock-like structure. For
a few cases, \cite{ontiveros09} were able to estimate the direction of
propagation of the shock.

This paper provides us with an opportunity to review the status of
CME-driven shocks in coronagraph images, and explain how these shocks
can be recognized in the observations and what parameters can be safely
extracted from calibrated images. We also present preliminary
observations of waves from the STEREO mission that can provide 3D
information about the shape, direction of the shock and better
estimates of the density compression ratio. 

\section{Where are the Shocks in Coronagraph Images?}
Our experience with coronagraph image analysis suggests that the lack
of unambiguous shock detections in the past is likely a result of the
lower contrast, smaller field of view, and reduced image cadence of
previous coronagraphs. The use of CCD imagers in the LASCO instruments
and the nearly continuous solar monitoring over 30 R$_{sun}$ enable
the detection of much fainter and finer structures in the images and
allows a better understanding of the observed scene; namely, the decoupling of
CME-related from other coronal structures. 

As we will argue, shock (and more generally wave)-induced density
enhancements are rather ubiquitous in the data but it is the use of
calibrated images and a familiarity with the morphology of these shock
signatures that allows an observer to distinguish them over the
plethora of ejecta and streamer material that may lie along a given
LOS. We demonstrate this via two, rather typical, shock morphologies
that we have been able to identify in the LASCO data.
\begin{figure}
  \includegraphics[height=.2\textheight]{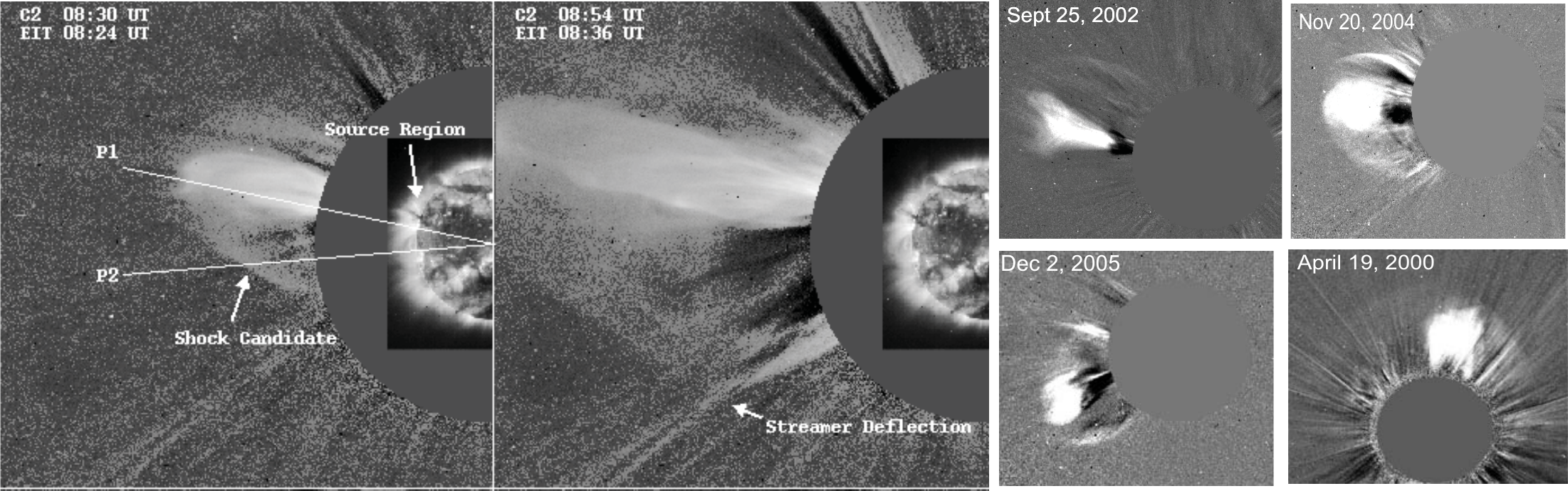}
  \caption{Examples of white light shock signatures with the bowshock
    morphology in LASCO images. Left panels: The April 2, 1999 event
    from Vourlidas et al (2003) showing the shock and streamer
    deflection. The lines labeled P1 and P2 mark the locations where
    density profiles were compared between observations and
    models. Right panels: Four more examples of bowshock-like
    events.}\label{fig:bowshock}
\end{figure}

\subsection{Bowshock Morphology}
In this case, the images show a density enhancement propagating from
the CME front along the CME flanks in a more or less straight line. It
looks very similar to the waves emanating for the bow of a ship or of
drawings of the bowshock of the terrestrial magnetosphere. The
obvious similarities to fluid shocks made these features prime
candidates for shock analysis and unsurprisingly were the first ones to
be verified as shocks \cite{vou03}. The two right panels in
Fig.~\ref{fig:bowshock} shows a very good example of bowshock
morphology taken from \cite{vou03}. The shock is driven by a narrow
ejection associated with a surge in the low corona. From our rather
limited search in the LASCO database, it appears that
this type of shock morphology is associated almost exclusively with
narrow CMEs as can been seen in the other examples (four right panels,
Fig.~\ref{fig:bowshock}).  

\subsection{``Double Front'' Morphology}
CME-driven shocks can be recognized in the images by another
characteristic morphology; a bright sharp loop-like feature proceeded
by a much fainter front. Faint emission fills in the space between the
two fronts. We use the term ``double front'' as a shorthand for referring
to this morphology. The fainter front is usually difficult to detect in direct
or even running difference images without some contrast enhancement
procedure. These features are easily detected in calibrated images
when a preevent image is subtracted. 
\begin{figure}
  \includegraphics[height=.3\textheight]{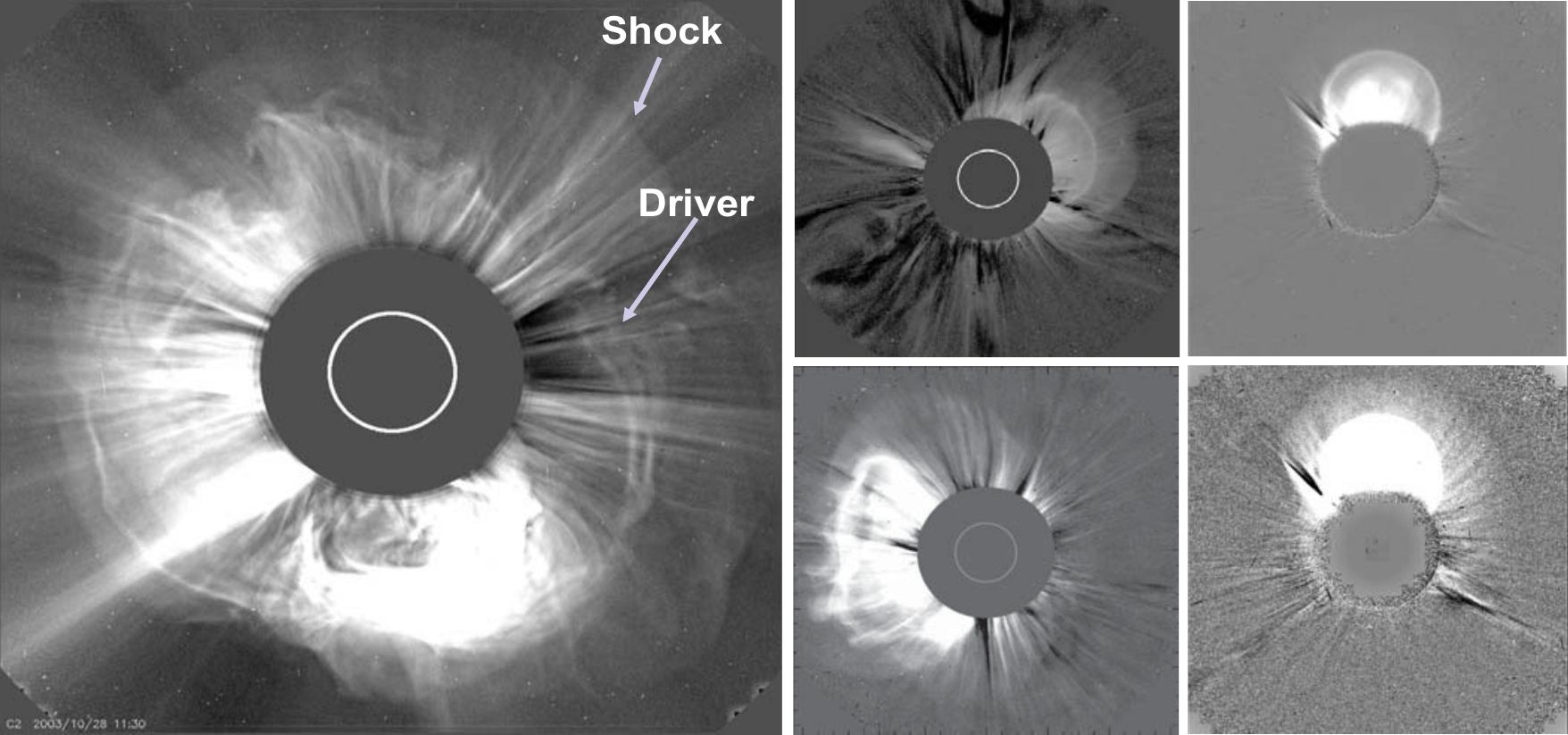}
  \caption{Examples of white light shocks with the 'double front'
    morphology in LASCO images. Right panel: The October 28, 2003 halo
    CME showing the shock and driver. Middle panels: Other examples of
    the ``double front'' morphology. Right panels: The same event can
    appear as a normal 3-part CME (top) or a ``double front'' (bottom)
    depending on the contrast (see text for discussion).} \label{fig:faint}
\end{figure}
Examples of ``double front'' events are shown in
Fig.~\ref{fig:faint}. This morphology can be easily understood when
one realizes that most CMEs are results of fluxrope eruptions
\cite{thernisien09}. Then, the bright loop is coronal plasma piled up
at the top of the erupting fluxrope and the fainter front is the shock
driven by the fluxrope. The fainter emission in-between the two front
is just a result of integrating through the larger shock structure
along the LOS. The faint front has been verified as a shock for one of
the events (October 28, 2003) by \cite{manch08} using an approach
similar to \cite{vou03}. We are very confident that this morphology
is a robust indicator of a shock. Such morphologies have been seen in
3D MHD models \cite{lynch04} but have not been recognized in the images
until now. This morphology is clearly associated with wide (partial or
full halo) and fast CMEs, and with so-called 3-part structure CMEs. It
appears then that the latter nomenclature is clearly a misnomer resulting from
lower contrast of past observations (right panels
of Fig.~\ref{fig:faint}). The implications of this discovery and a
detailed study of these events with comparisons to models will be
discussed in an upcoming paper \cite{vou09}.

\section{Extraction of the Shock Physical Parameters}
\begin{figure}
  \includegraphics[height=.25\textheight]{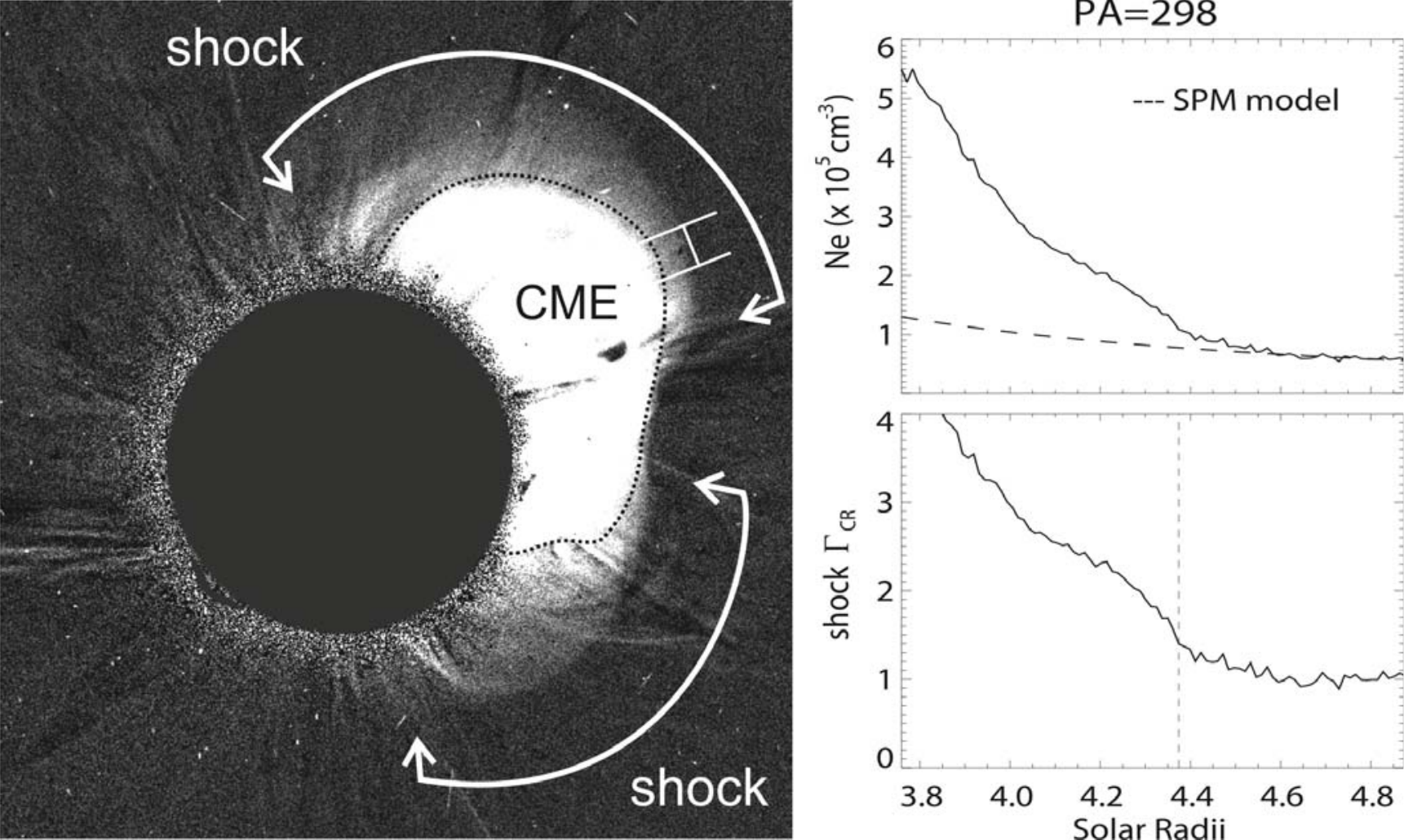}
  \caption{Density profile across the front of white light shocks. The
    shock, CME, and the location of the density profile are marked on
    the left panel. The excess density and compression ratio are shown
    on the right panel. The dashed line in the top panel is the
    assumed background density. The vertical dotted line in the bottom
    panel marks the location of the shock front.} \label{fig:shocks}
\end{figure}

The above morphologies are just two of the most easily recognized
signatures of shock in the images. Shock fronts exhibit much more
variability in the images. They can be continuous or appear over a
small range of position angles. Sometimes they are detected ahead of
the CME nose and sometimes not. Many times, no such fronts are seen
ahead of fast CMEs but they can been seen in their flanks. Other times
there is no evidence for a faint front anywhere around the image. In
the majority of these cases, though, the CMEs propagate in the wake of
another CME which disturbs the environment. Generally speaking almost
all fast CMEs exhibit a faint front somewhere ahead of the main
event. When such front is detected, we can extract some physical
parameters from it, such as density compression ratio, speed, and even
direction, as was shown by \cite{ontiveros09} (Fig.~\ref{fig:shocks}).
\subsubsection{Compression Ratios}
To derive a density compression ratio, we first estimate the density
at the shock and the preevent density. The latter is usually
determined by the inversion of partial polarized (pB) images taken
near the event. For the LASCO images, normally only one pB sequence is
taken for each of the coronagraphs (C2, C3). Alternatively, a standard
density model such as the SPM \cite{saito77} can be used. The shock
density can be estimated by the measured excess mass (or number of
electrons) at the shock front with the assumption of the (unknown)
depth along the LOS. Results are shown in Fig.~\ref{fig:shocks}

\begin{figure}
  \includegraphics[height=.25\textheight]{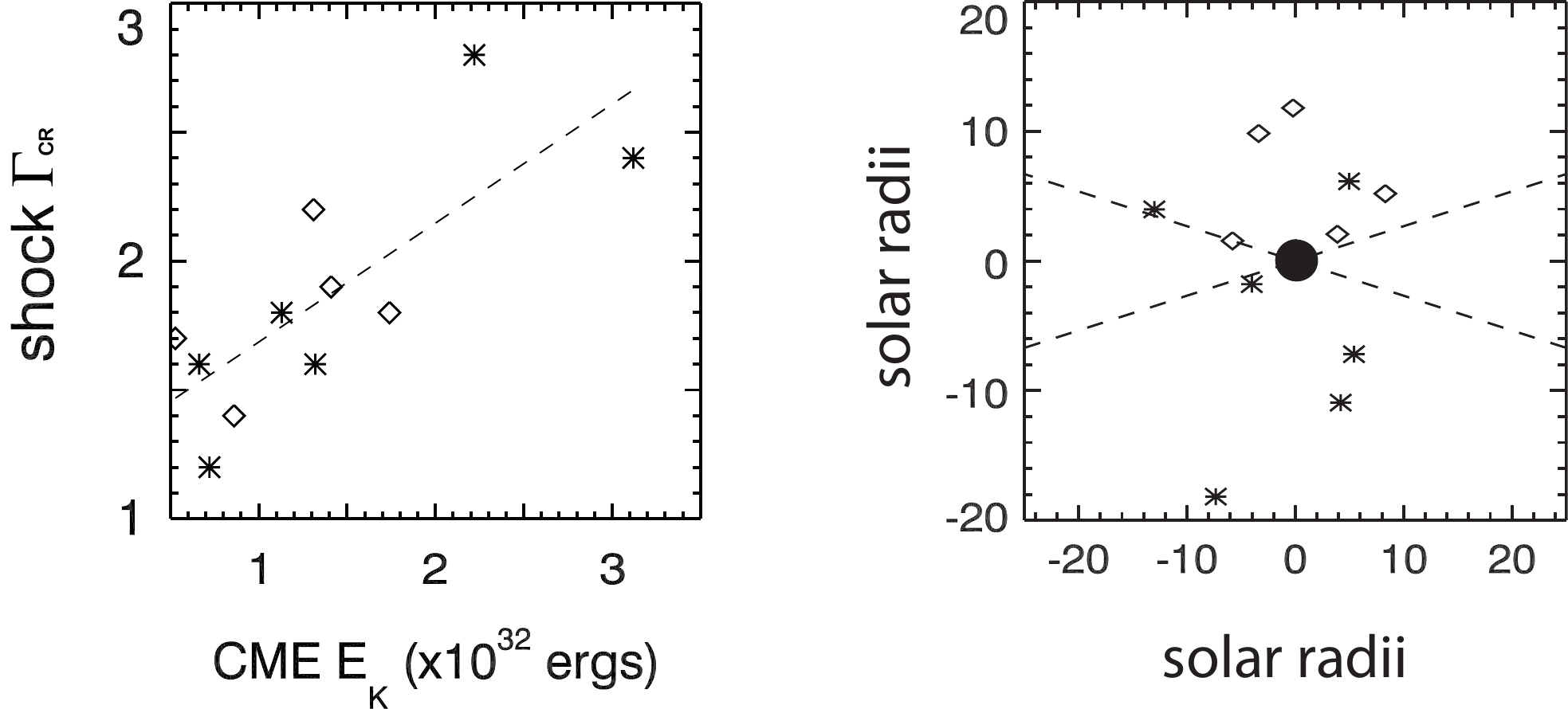}
  \caption{Analysis of shock parameters in \cite{ontiveros09}. Left
    panel: The density jump at the shock front shows a good
    correlation with the CME kinetic energy. Right panel: Plane of sky
    locations of the best shock signatures. The plot reveals that shock
    fronts are more visible when projected away from the streamer belt.} \label{fig:prof}
\end{figure}

For the events studied by \cite{ontiveros09}, the resulting density
jump is between 1.1 and 2.8 which is consistent with in-situ
measurements and shows a reasonable correlation with the CME kinetic
energy as expected (Fig.~\ref{fig:prof}). Because the
emission is optically thin, all white light measurements are subject
to projection effects. The projection effects tend to confuse the
shock with other intervening structures along the LOS
(Fig~\ref{fig:prof}) and smooth the density jump in the images
compared to the sharp jumps observed with in-situ
instrumentation. \cite{ontiveros09} demonstrated that this is indeed
the case by fitting the observed shock shape with a 3-dimensional
geometric model of a bowshock shell with a width of 0.3 R$_{sun}$ and
calculating the integrated emission from it. The resulting modeled
density profiles fit very well the observed ones. This is encouraging
and suggests that we may be able to extract 'true' density compression
ratios from the images at heliocentric heights unreachable by in-situ
probes but important for understanding particle acceleration.

\subsubsection{Shock Shape and Direction}
The geometric modeling of the shock provides another very important
parameter: the direction of the shock nose. For the three events
analyzed by \cite{ontiveros09} it was found that the shock nose was
within 30$^\circ$ of the radial from the likely source
region. Although the reliability of this method needs to be validated
with more events, it suggests that it could be possible to estimate
the direction of a shock from single point coronagraph observations
early in its evolution. 

\subsubsection{3-Dimensional  Shock Measurements}
Shock analysis will greatly benefit from the operation of the twin
SECCHI coronagraphs suite \cite{howard08} aboard the
\textsl{STEREO\/} mission \cite{kaiser08}. The ability to image the
CME and its shock from two viewpoints simultaneously enables the
fitting of the 3D shape of the shock with much more fidelity that it
is possible from a single viewpoint. It also allows the easy
separation of the shock from the driver and from other coronal
features. 

\begin{figure}
  \includegraphics[height=.3\textheight]{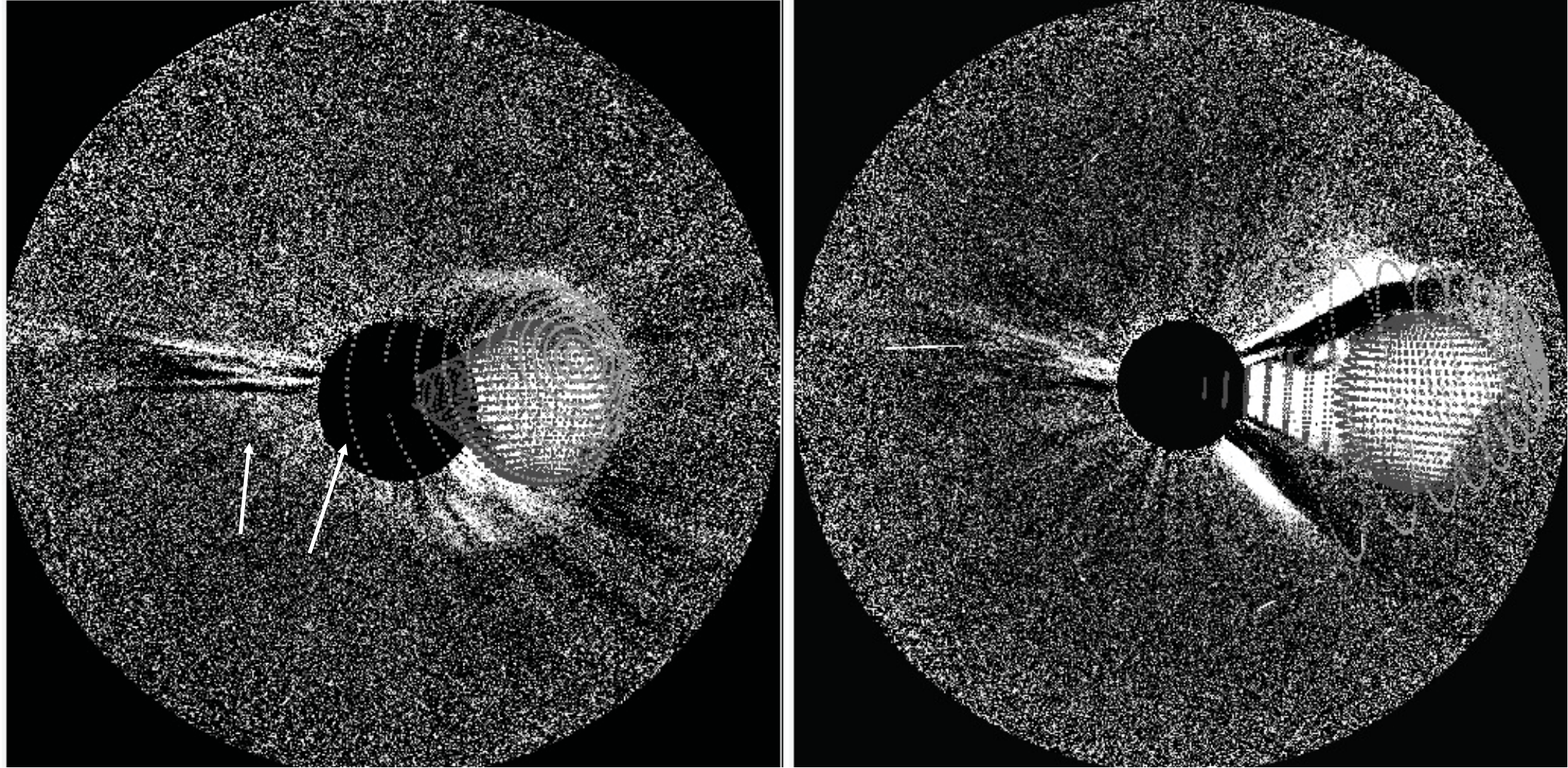}
  \caption{Example of the use of simultaneous SECCHI/COR2 observations for
    3D forward modeling of a CME (dark frame) and its driven wave
    (light grey frame). The COR2-B (COR2-A) image is the left
    (right) panel. The arrows show that the eastern extent of the model
  does not agree with the observed wave and suggests that the shock
  does not propagate isotropically in the corona.} \label{fig:stereo}
\end{figure}

Because of the protracted solar activity minimum, there have been only
a couple of shocks detected by SECCHI. Analysis of the December 31,
2007 event has shown that the shock direction and its propagation
along the flanks of the streamer can be reconstructed in 3D
\cite{ontiveros09b}. Because of the higher sensitivity of the SECCHI
coronagraphs it is now possible to reconstruct the waves driven by the
slower CMEs which are common during solar
minimum. Fig.~\ref{fig:stereo} shows an example of simultaneous
forward modeling of the CME-driven wave (light grey frame) and the CME
itself (dark frame). The model shows that the CME propagates at
116$^\circ$ west of the Sun-Earth line and the shock at
121$^\circ$. It suggests that the shock and CME direction
diverge. This is supported by observations at larger distances (not
shown here). Another interesting result is the mismatch between the
eastern extent of the model compared and the actual observations that
show a much larger extent. It is obvious that a symmetric bowshock is
not a good model for the actual shock which appears to propagate
anisotropically in the corona. This behavior is expected since the
morphology of the large scale coronal field will affect the propagation
of an alfvenic wave by guiding it through the minimum of the local
Alfv\'{e}n speed.
  
\section{Conclusions}
We cannot fully cover, in the small space allocated here, the
implications from the recent advances in the detection and measurement
of white light shocks. We give only a general overview of the
results to demonstrate two things: (1) that
CME-driven shocks can be easily detected in calibrated coronagraph
images, and (2) that useful parameters of the shock properties can be
extracted from these images. Plenty of work remains to be done but we
have established some important facts:
\begin{itemize}
\item The observed shock intensities and shapes are consistent with an
  LOS integrates emission from a thin shell resembling a bowshock.
\item The density compression ratios are less than 3, in agreement
  with in-situ measurements.
\item The visibility of the shock depends strongly on the structures
  along the LOS. A preceding event may disturb the corona
  sufficiently to hamper detection of the faint shock signatures
\item It is relatively straightforward to derive the direction of the
  shock from two viewpoint image. It may be possible to estimate the
  direction of the shock from single viewpoint images.
\end{itemize}
The ability to identify the shock structures, including the associated
streamer deflections, is an important step towards the ultimate goal
of the proper interpretation of coronagraph images. It will result in
more accurate measurements of CMEs and their properties by enabling
the observers to separate the ejecta from other structures in the
images and by permitting a better connection with features seen in
other regimes (e.g., EUV images).

Finally, the analysis of white light shocks will greatly benefit the
understanding of shock and particle acceleration physics when combined
with off-limb spectroscopy and in-situ particle measurements like
those planned for the upcoming Solar Orbiter mission. 


\begin{theacknowledgments}
  Part of this work was funded by the LWS TR\&T grant NNH06D85I. SOHO
  is an international collaboration between NASA and ESA. LASCO was
  constructed by a consortium of institutions: NRL (Washington, DC,
  USA), MPS (Katlenburg- Lindau, Germany), LAM (Marseille, France) and
  Univ.of Birmingham (Birmingham, UK). The SECCHI data are produced by
  an international consortium of the NRL, LMSAL and NASA GSFC (USA),
  RAL and Univ. Bham (UK), MPS (Germany), CSL (Belgium), IOTA and IAS
  (France).

\end{theacknowledgments}

\bibliographystyle{aipproc}   

\end{document}